\documentclass[12pt]{iopart}
\usepackage[dvips]{graphicx}
\usepackage{bm}                      % bold math
\usepackage{amssymb}                 % ams symbols

\begin{document}

\def\bea{\begin{eqnarray}}
\def\eea{\end{eqnarray}}
\def\be{\begin{equation}}
\def\ee{\end{equation}}
\def\rra{\right\rangle}
\def\lla{\left\langle}
\def\sig{\sigma}
\def\eps{\epsilon}
\def\sgm{\Sigma^-}
\def\la{\Lambda}
\def\pv{\bm{p}}
\def\kv{\bm{k}}
\def\tv{\bm{\tau}}
\def\sv{\bm{\sigma}}
\def\rv{\bm{r}}
\def\sdot{\!\cdot\!}
\def\ns{neutron star}
\def\bc{B=90\;\rm MeV\!/fm^3}
\def\ms{M_\odot}

\title{The Equation of State of Dense Matter: from Nuclear Collisions to
Neutron Stars}

\author{G. F. Burgio}

\address{Istituto Nazionale di Fisica Nucleare, Sez. di Catania,
Via S. Sofia 64, 95123 Catania, Italy } 

\ead{fiorella.burgio@ct.infn.it}

\begin{abstract}
The Equation of State (EoS) of dense matter represents a central issue in the 
study of compact astrophysical objects and 
heavy ion reactions at intermediate and relativistic energies. 
We have derived a nuclear EoS with nucleons and hyperons
within the Brueckner-Hartree-Fock approach, and joined it 
with quark matter EoS. For that, we have employed the MIT bag model, as well as
the Nambu--Jona-Lasinio (NJL) and the Color Dielectric (CD) models, and
found that the NS maximum masses are not larger than 1.7 solar masses.
A comparison with available data supports the idea
that dense matter EoS should be soft at low density and quite stiff at high 
density.

\end{abstract}
%\maketitle

\section{Introduction}

In the last few years, the study of the equation
of state  of nuclear matter has stimulated an intense theoretical activity.
The interest for the nuclear EoS lies, to a large extent, 
in the study of compact objects, i.e., supernovae and neutron stars. 
In particular, the structure of a neutron star is very sensitive 
to the compressibility and the symmetry energy. For that, several 
phenomenological and microscopic models of the EoS have been developed.
The former models include nonrelativistic
mean field theory based on Skyrme interactions \cite{bon} and
relativistic mean field theory based on meson-exchange
interactions (Walecka model) \cite{wal}. 
The latter ones include nonrelativistic
Brueckner-Hartree-Fock (BHF) theory \cite{bal} and its relativistic
counterpart, the Dirac-Brueckner (DB) theory \cite{dbhf,tueb}, and the
nonrelativistic variational approach also corrected by
relativistic effects \cite{pan}. 
In these approaches the parameters
of the interaction are fixed by the experimental nucleon-nucleon (NN)
and/or nucleon-meson scattering data.

One of the most advanced microscopic approaches to the EoS of 
nuclear matter is the Brueckner theory. 
In the recent years, it has made a rapid progress in several aspects:
(i) The convergence of the
Brueckner-Bethe-Goldstone (BBG) expansion has been firmly
established \cite{thl}. 
(ii) The addition of three-body forces (TBF) permitted to
the agreement with the empirical
saturation properties \cite{gra,lom}.
(iii) The extension of the BHF approach has to the description of
 nuclear matter containing also 
hyperons \cite{hypmat}, thus leading to 
a more realistic modeling of neutron stars \cite{hypns,mmy}.

In the present paper we review these issues and
present our results for neutron star structure 
based on the resulting EoS of dense hadronic matter,
also supplemented by an eventual transition to quark matter
at high density. A comparison with available experimental data from heavy ion 
collisions and neutron stars' observations will be discussed. 

%-----------------------------------------------------------------------
\section{The Equation of State from the BBG approach}

The Brueckner--Bethe--Goldstone (BBG) theory \cite{bal} is based on a linked
 cluster 
expansion of the energy per nucleon of nuclear matter. 
The basic ingredient in this many--body approach is the Brueckner reaction 
matrix $G$, which is the solution of the  Bethe--Goldstone equation 

\begin{equation}
G(\rho;\omega) = v  + v \sum_{k_a k_b} {{|k_a k_b\rangle  Q  \langle k_a k_b|}
  \over {\omega - e(k_a) - e(k_b) }} G(\rho;\omega), 
\end{equation}                                                           
\noindent
where $v$ is the bare nucleon-nucleon (NN) interaction, $\rho$ is the nucleon 
number density, $\omega$  is the  starting energy, and   
$|k_a k_b\rangle Q \langle k_a k_b|$  is  the Pauli operator.  
$e(k) = e(k;\rho) = {{\hbar^2}\over {2m}}k^2 + U(k;\rho)$
is the single particle energy, and the  
Brueckner--Hartree--Fock (BHF) approximation for the single particle potential
$U(k;\rho)$ reads
$U(k;\rho) = \sum _{k'\leq k_F} \langle k k'|G(\rho; e(k)+e(k'))|k k'\rangle_a $
(the subscript ``{\it a}'' indicates antisymmetrization of the 
matrix element). In the BHF approximation the energy per nucleon is
 
\begin{equation}
{E \over{A}}  =  
          {{3}\over{5}}{{\hbar^2~k_F^2}\over {2m}} + D_{\rm BHF} ~~,
D_{\rm BHF}(n) = {{1}\over{2A}}  
\sum_{k,k'\leq k_F} \langle k k'|G(\rho; e(k)+e(k'))|k k'\rangle_a 
\end{equation}
\noindent
In this scheme, the only input quantity we need is the bare NN interaction
$v$ in the Bethe-Goldstone equation (1). In this sense the BBG 
approach can be considered as a microscopic approach. However, 
it is well known that two-body forces are not able to
explain some nuclear properties (e.g., binding energy of light nuclei, and 
saturation point of nuclear matter), and three-body forces (TBF)
have to be introduced.
In the framework of the Brueckner theory, a rigorous treatment of TBF
would require the solution of the Bethe-Faddeev equation,
describing the dynamics of three bodies embedded in the nuclear matter. 
In practice a much simpler approach is employed, namely
the TBF is reduced to an effective,
density-dependent, two-body force by averaging over the third
nucleon in the medium, taking account of the nucleon-nucleon
correlations. This effective two-body
force is added to the bare two-body force and recalculated at each
step of the iterative procedure.

Both phenomenological and microscopic TBF have been used in the BHF approach.
The phenomenological TBF is widely used in the literature, in particular
the Urbana IX TBF \cite{uix} for variational calculations of finite nuclei and 
nuclear matter \cite{pan}, and contains a two-pion
exchange potential, which is attractive at low density, and
a phenomenological repulsive term, more effective at high density.
The microscopic TBF is based on 
meson-exchange mechanisms accompanied by the excitation of
nucleonic resonances \cite{gra,lom}, and produces
a remarkable improvement of the saturation properties
of nuclear matter \cite{lom}. 
\begin{figure} [ht]
\centerline{
\includegraphics[width=0.6\textwidth]{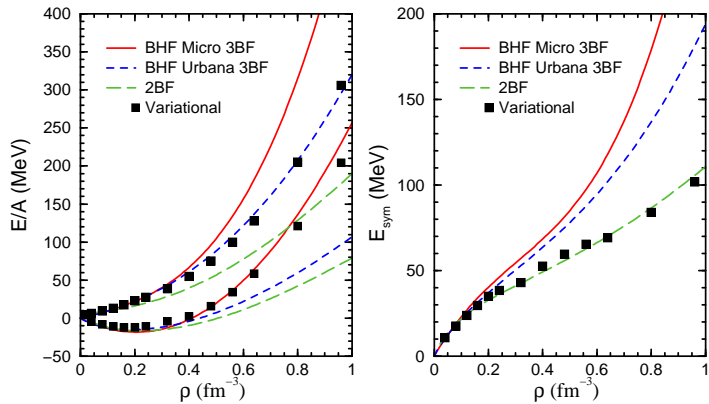}}
\caption{In the left panel, symmetric matter (lower curves) and pure 
 neutron matter 
(upper curves) EoS, calculated within the BBG approach, are shown.
Variational many-body calculations are also displayed (full squares).
The symmetry energy is shown in the right panel.}
    \label{f:Fig1}
\end{figure}

Let us now compare the EoS predicted by the BHF
approximation with the same two-body force 
(Argonne $v_{18}$ \cite{v18}) and different TBF \cite{zhou}.
In the left panel of Fig.~\ref{f:Fig1} we display the EoS 
both for symmetric matter
(lower curves) and pure neutron matter (upper curves).
We show results obtained for several cases, i.e.,
i) only two-body forces are included (long dashed lines),
ii) TBF treated within the phenomenological Urbana IX model 
(dashed lines), and the microscopic meson-exchange approach 
(solid lines). For completeness, we also show results obtained with 
variational calculations (full squares) \cite{pan}.
We notice that the EoS for symmetric matter with TBF
reproduces the correct nuclear matter saturation point in all approaches.
Moreover, up to a density of $\rho \approx 0.4\;{\rm fm}^{-3}$
the BHF EoS calculated with  TBF 
are in fair agreement with the variational calculations,
whereas at higher density the microscopic TBF turns out to be the most
repulsive.
In all cases, the incompressibility at saturation is compatible
with the values extracted from phenomenology, i.e.,
$K \approx 210\;\rm MeV$.
In the right panel of Fig.~\ref{f:Fig1} we display the symmetry energy as a
function of the nucleon density $\rho$.
Within the BHF approach, the symmetry energy has been calculated 
within the so-called  ``parabolic approximation'' for the binding
energy of nuclear matter with arbitrary proton fraction \cite{bom}.
We observe results in agreement with the characteristics of the
EoS shown in the left panel, namely, the stiffest EoS
yields larger symmetry energies compared to the ones obtained with the
Urbana phenomenological TBF and the variational calculations.
This leads to a different proton fraction in beta-stable nuclear matter.
We notice that the symmetry energy calculated (with or without TBF) 
at the saturation point yields a value $E_{\rm sym}\approx 30\;\rm MeV$,
compatible with nuclear phenomenology.

\begin{figure} [ht]
\centerline{
\includegraphics[angle=270,width=0.42\textwidth]{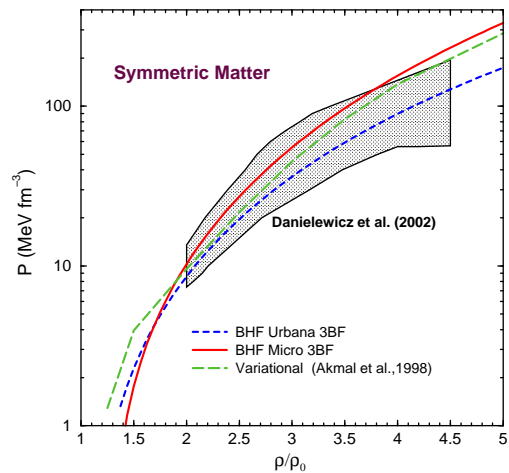}}
\caption{Different EoS are compared with the phenomenological constraint
    extracted by Danielewicz et al. \cite{daniel} (shaded area).
    Solid (dashed) line: BHF EoS with microscopic (phenomenological)
    TBF. Long dashed line : variational  EoS. }
    \label{f:Fig2}
\end{figure}

In the last few years it became popular to compare the various microscopic and
phenomenological EoS with the allowed region in the pressure--density plane, as
determined by Danielewicz et al. \cite{daniel}. In that paper the authors
consider both the in--plane transverse flow and the elliptic flow
measured  in different experiments on $ Au + Au $ collisions at
energies between 0.2 and 10 GeV/A. From the data, Danielewicz et al. 
could  estimate the pressure for symmetric matter. 
In Fig.~\ref{f:Fig2} the set of microscopic EoS 
discussed is displayed along with the allowed pressure region 
(shaded area). Both the EoS derived from BHF with Urbana IX TBF
and the variational one are in
agreement with the phenomenological analysis, while the BHF EoS 
with microscopic TBF turns out to be only marginally
compatible, since at higher density it becomes too stiff and
definitely falls outside the allowed region. Additional analyses of flow data,
as reported by the FOPI Collaboration \cite{stoi}, and subthreshold $K^+$ 
production \cite{kaos} confirm a soft equation of state in the same density 
range (see C. Fuchs contribution to this conference).

%-------------------------------------------------------------------------------
\section{Hyperons in nuclear matter}

While at moderate densities $\rho \approx \rho_0$ the matter inside 
a neutron star consists only of nucleons and leptons, 
at higher densities several other
species of particles may appear due to the fast rise of the baryon 
chemical potentials with density. 
Among these new particles are strange baryons, namely, 
the $\Lambda$, $\Sigma$, and $\Xi$ hyperons. 
Due to its negative charge, the $\Sigma^-$ hyperon is the 
first strange baryon expected to appear with increasing density in the 
reaction $n+n \rightarrow p+\Sigma^-$,
in spite of its substantially larger mass compared to the neutral 
$\Lambda$ hyperon ($M_{\Sigma^-}=1197\;{\rm MeV}, M_\Lambda=1116\;{\rm MeV}$).
Other species might appear in stellar matter,
like $\Delta$ isobars along with pion and kaon condensates.

We have generalized the study of the nuclear EoS
with the inclusion of the $\Sigma^-$ and $\Lambda$ hyperons in the BHF
many-body approach.
To this purpose, one requires in principle 
nucleon-hyperon (NH) and hyperon-hyperon (HH) potentials. 
In our work we use the Nijmegen soft-core NH potential \cite{nsc89},
that is well adapted to the existing experimental NH scattering data.
Unfortunately, up to date no HH scattering data exist 
and therefore no reliable HH potentials are available.
Hence we neglected HH potentials in our BHF calculations \cite{hypns}.
Nevertheless, the importance of HH potentials should be minor 
as long as the hyperonic partial densities remain limited.

\begin{figure} [ht]
\centerline{
\includegraphics[angle=270,width=0.7\textwidth]{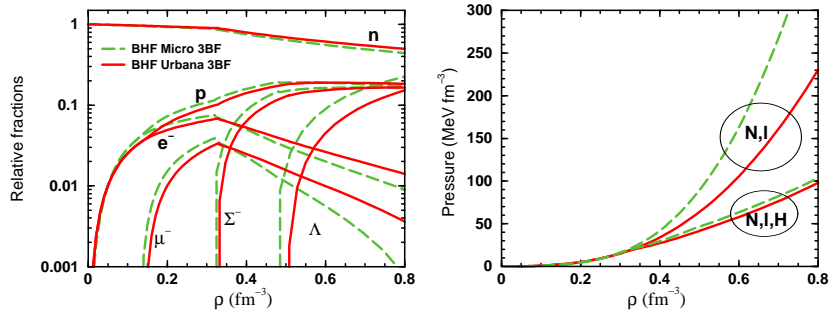}}
\caption{The particle concentrations (left panel) are shown as function of the 
         baryon density. Long dashed curves are calculations performed with 
         the microscopic TBF, whereas solid lines represent the Urbana TBF 
         calculations. In the right panel the corresponding EoS are shown.}
    \label{f:Fig3}
\end{figure}

In Fig.~\ref{f:Fig3} (left panel) we show the chemical composition of 
the resulting
beta-stable and asymmetric nuclear matter containing hyperons.
We observe rather low hyperon onset densities of about 2-3 times 
normal nuclear matter density
for the appearance of the $\sgm$ and $\la$ hyperons, almost independently
on the adopted TBF.
Moreover, an almost equal percentage of nucleons and hyperons are 
present in the stellar core at high densities. 
A strong deleptonization of matter takes place, 
and this can have far reaching consequences for the 
onset of kaon condensation \cite{kaon}.
The resulting EoS is displayed in the right panel of Fig.~\ref{f:Fig3}.
The upper curves show the EoS when stellar matter is composed 
only of nucleons and leptons, whereas the lower curves show 
calculations with nucleons and hyperons. We notice that the inclusion 
of hyperons produces a much softer EoS,
no matter the TBF adopted in the nucleonic sector.
These remarkable results are due to the inclusion of hyperons as additional 
degrees of freedom,
and we do not expect substantial changes when introducing refinements
of the theoretical framework, such as 
hyperon-hyperon potentials, hyperonic TBF, relativistic corrections, etc. 
\begin{figure} [ht]
 \centerline{
\includegraphics[angle=270,width=0.35\textwidth]{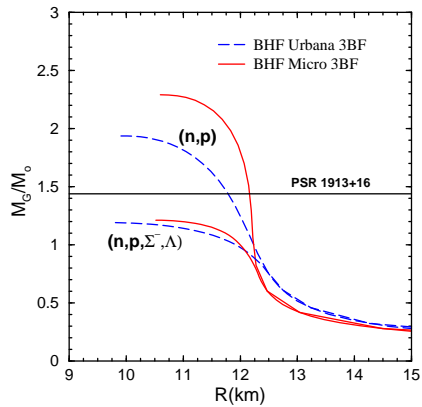}}
   \caption{The mass-radius relation is plotted for EoS without hyperons 
(upper curves), and with hyperons (lower curves). See text for details.}
    \label{f:Fig4}
\end{figure}

The consequences for the structure of the neutron stars are illustrated  
in Fig.~\ref{f:Fig4}, where we display the resulting neutron star 
mass-radius curves, obtained solving the Tolman-Oppenheimer-Volkoff 
equations \cite{ns}.
We notice that the BHF EoS calculated with the microscopic TBF produces
the largest gravitational masses , with the maximum mass of the order of
2.3 $\ms$, whereas the phenomenological TBF yields a maximum mass of
about 1.9 $\ms$.
In the latter case, neutron stars are characterized by smaller radii and
larger central densities, i.e.,
the Urbana TBF produce more compact stellar objects.
One should notice that, although different TBF still yield
quite different maximum masses,
the presence of hyperons equalizes the results, 
leading now to a maximum mass of less than 1.3 solar masses
for all the nuclear TBF. This result is in contradiction with the 
measured  value of the Hulse-Taylor pulsar mass, PSR1913+16, 
which amounts to 1.44 $M_\odot$.
The only remaining possibility in order to reach significantly larger 
maximum masses appears to be the transition to another phase of 
dense (quark) matter inside the star. 
This is indeed a reasonable assumption, since already geometrically 
the concept of distinguishable baryons breaks down
at the densities encountered in the interior of a neutron star.
This will be discussed in the following.

%-------------------------------------------------------------------------------
\section{Quark matter}

The results obtained with a purely hadronic EoS call for an estimate of
the effects due to the hypothetical presence of quark matter in the interior
of the neutron star.
Unfortunately, the current theoretical description of quark matter 
is burdened with large uncertainties, seriously limiting the 
predictive power of any theoretical approach at high baryonic density. 
For the time being we can therefore only resort
to phenomenological models for the quark matter EoS,
and try to constrain them as well as possible 
by the few experimental information on high-density baryonic matter.

One of these constraints is the phenomenological
observation that in heavy ion collisions at intermediate energies
($10\;{\rm MeV}/A \lesssim E/A \lesssim 200\;{\rm MeV}/A$) 
no evidence for a transition to a quark-gluon plasma has been found
up to about 3$\rho_0$.
We have taken this constraint in due consideration, 
and used an extended MIT bag model \cite{chodos}
(including the possibility of a density dependent bag ``constant'') 
and the color dielectric model \cite{pir}, 
both compatible with this condition \cite{cdm}. For completeness, we have also
used the Nambu--Jona-Lasinio model \cite{njl}.

In order to study the hadron-quark phase transition in neutron stars, 
we have performed the Maxwell construction, so demanding
a sharp phase transition. 
We have found that the phase transition in the extended MIT bag model 
takes place at a large baryon density, $\rho \approx 0.6\;\rm fm^{-3}$,
and at larger baryon density in the NJL model \cite{njl}. 
On the contrary, the transition density in the CD model is  
$\rho \approx 0.05\;\rm fm^{-3}$.
This implies a large difference in the structure of hybrid stars.
In fact, whereas stars built with the CD model have at most a mixed phase
at low density and a pure quark core at higher density, the ones obtained
with the MIT bag model contain a hadronic phase, followed by a mixed phase
and a pure quark interior. The scenario is again different 
within the Nambu-Jona--Lasinio model, where at most a mixed phase
is present, but no pure quark phase. 
\begin{figure} [ht]
 \centerline{
\includegraphics[angle=270,width=0.4\textwidth]{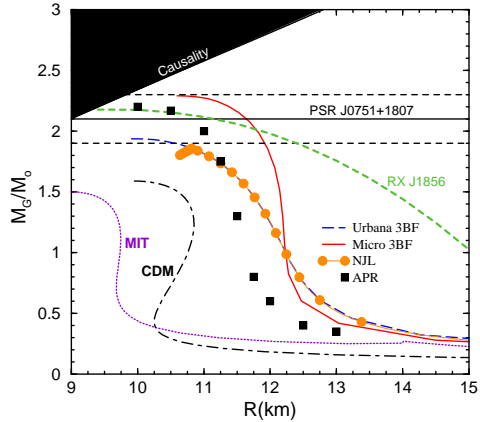}}
   \caption{The mass-radius relation is shown for the several cases discussed 
in the text, along with some observational constraints.}
    \label{f:Fig5}
\end{figure}
The final result for the structure of hybrid neutron stars 
is shown in Fig.~\ref{f:Fig5}, displaying the mass-radius relation.
It is evident that the most striking effect of the inclusion of quark
matter is the increase of the maximum mass with respect to the case with
hyperons, now reaching about 
$1.5\;\ms$.
At the same time, the typical neutron star radius is reduced by about
3 km to typically 9 km. 
Hybrid neutron stars are thus more compact than purely hadronic ones
and their central energy density is larger.
In Fig.~\ref{f:Fig5} we also display some observational constraints. 
The first one demands that any reliable 
EoS should be able to reproduce the recently reported high pulsar mass of
$\rm 2.1 \pm 0.2~M_\odot$ for PSR J0751+1807 \cite{nice}. 
Extending this value even to 
2$\sigma$ confidence level $(_{-0.5}^{+0.4}M_\odot)$ means that masses of at 
least $\rm 1.6~M_\odot$ have to be allowed. The other constrain comes from a
recent analysis of the thermal radiation of the isolated pulsar RX J1856 which
determines a lower bound for its mass-radius relation that implies a rather 
stiff EoS \cite{trum}.  Both constraints indicate that the EoS should be 
rather stiff at high density. Moreover, if quark matter is 
present in the neutron stars' interiors, this would require additional 
repulsive contributions in the quark matter EoS. 

%-----------------------------------------------------------------------------
\section{Conclusions}

In this paper we reported the theoretical description of nuclear
matter in the BHF approach, 
with the application to neutron star structure calculation.
We pointed out the important role of TBF at high density,
which is, however, strongly compensated by the inclusion of hyperons.
The resulting hadronic neutron star configurations have maximum masses 
of less than $1.4\;\ms$, and the presence of quark matter inside the star 
is required in order to reach larger values.

Concerning the treatment of quark matter, we have
joined the corresponding EoS with the hadronic 
one, and reached maximum masses of about $1.7\;\ms$.
The value of the maximum mass of neutron stars obtained according to
our analysis appears rather robust with respect to the uncertainties
of the nuclear and the quark matter EoS.
Therefore, the experimental observation of a very heavy
($M \gtrsim 1.7\; \ms$) neutron star would suggest that 
serious problems are present for the current theoretical modelling
of the high-density phase of nuclear matter.
In any case, one can expect a well defined hint on the
high-density nuclear matter EoS.

%%%%%%%%%%%%%%%%%%%%%%%%%%%%%%%%%%%%%%%%%%%%%%%%%%%%%%%%%%%%%%%%

\end{document}